\documentclass[a4paper,usenatbib]{mnras}

\usepackage[T1]{fontenc}
\usepackage{ae,aecompl}

\usepackage{graphicx}	
\usepackage{amsmath}	
\usepackage{amssymb}	
\usepackage{hyperref}

\usepackage{bm}
\usepackage{epsfig}
\usepackage{dcolumn}
\usepackage{xcolor}
\usepackage{amssymb}
\usepackage{color}
\usepackage{epsf}
\usepackage{float}
\usepackage{latexsym}
\usepackage[utf8]{inputenc}
\usepackage{url}
\usepackage{ulem}

\title[Scale-dependent hemispherical asymmetry]
{Testing the scale-dependent hemispherical asymmetry with the 21-cm power spectrum from the epoch of reionization}
\author[B.~Li et al.]
{Botao Li,$^{1,2,3}$\thanks{E-mail: lbt@mail.ustc.edu.cn}
Zhaoting Chen,$^{1,2,3}$\thanks{E-mail: czt180@mail.ustc.edu.cn}
Yi-Fu Cai,$^{1,2,3}$\thanks{Corresponding author. E-mail: yifucai@ustc.edu.cn}
and Yi Mao$^{4}$\thanks{Corresponding author. E-mail: ymao@tsinghua.edu.cn} \\
$^{1}$ Department of Astronomy, School of Physical Sciences, University of Science and Technology of China, Hefei, Anhui 230026, China \\
$^{2}$ CAS Key Laboratory for Research in Galaxies and Cosmology, University of Science and Technology of China, Hefei, Anhui 230026, China \\
$^{3}$ School of Astronomy and Space Science, University of Science and Technology of China, Hefei, Anhui 230026, China\\
$^{4}$ Department of Astronomy and Tsinghua Center for Astrophysics, Tsinghua University, Beijing 100084, China
}

\date{Accepted 2019 June 6. Received 2019 June 6; in original form 2019 April 9.}

\pubyear{2019}

\begin{document}
\label{firstpage}
\pagerange{\pageref{firstpage}--\pageref{lastpage}}
\maketitle

\begin{abstract}

Hemispherical power asymmetry has emerged as a new challenge to cosmology in early universe. While the cosmic microwave background (CMB) measurements indicated the asymmetry amplitude $A \simeq 0.07$ at the CMB scale $k_{\rm CMB}\simeq 0.0045\,{\rm Mpc}^{-1}$, the high-redshift quasar observations found no significant deviation from statistical isotropy. This conflict can be reconciled in some scale-dependent asymmetry models. We put forward a new parameterization of scale-dependent asymmetric power spectrum, inspired by a multi-speed inflation model. The 21-cm power spectrum from the epoch of reionization can be used to constrain the scale-dependent hemispherical asymmetry. We demonstrate that an optimum, multi-frequency observation by the Square Kilometre Array (SKA) Phase 2 can impose a constraint on the amplitude of the power asymmetry anomaly at the level of $\Delta A \simeq 0.2$ at $0.056 \lesssim k_{\rm 21cm} \lesssim 0.15 \,{\rm Mpc}^{-1}$. This limit may be further improved by an order of magnitude as $\Delta A \simeq 0.01$ with a cosmic variance limited experiment such as the Omniscope.

\end{abstract}


\section{Introduction}

Common wisdom has it that there exists no preferred location or preferred direction on large scales in our universe, as dictated by the cosmological principle, namely, that the background of this universe is isotropic and homogeneous at cosmic distances. Tiny quantum fluctuations in density at early era can give rise to primordial inhomogeneities and seed the large-scale structure (LSS) that we see today \citep{Mukhanov:1990me}. In standard paradigms of the very early universe, such as the inflationary cosmology \citep{Guth:1980zm, Linde:1981mu, Albrecht:1982wi, Starobinsky:1980te, Fang:1980wi, Sato:1980yn}, these primordial density fluctuations result in a nearly scale-invariant power spectrum of curvature perturbation, which has been confirmed in high precision by cosmic microwave background (CMB) observations \citep{Bennett:2003bz, Hinshaw:2012aka, Ade:2015xua}.

However, some potential deviation from statistical isotropy in the temperature fluctuations has been indicated by CMB experiments for years, which is known as the CMB anomaly of hemispherical power asymmetry. This asymmetry, i.e. the dipolar modulation of the CMB temperature at cosmological scales, was first pointed out by the Wilkinson Microwave Anisotropy Probe (WMAP) satellite \citep{Eriksen:2003db, Hansen:2004vq, Eriksen:2007pc, Hoftuft:2009rq}, and further reported by the Planck team with about $3\sigma$ significance level \citep{Ade:2013nlj, Akrami:2014eta, Ade:2015hxq}. The amplitude of this asymmetry, which is often parametrized by a dimensionless parameter $A$, is found to be $A \sim 0.07$ and $\Delta A \sim 0.02$ \citep{Ade:2015hxq} at large scales (corresponding to small $\ell$), $\ell \lesssim 64$. However, at small scales, results from both Sloan Digital Sky Survey (SDSS) quasar sample observations \citep{Hirata:2009ar} and the CMB high $\ell$ power spectrum measurement \citep{Aiola:2015rqa} are consistent with statistical isotropy, i.e. the power asymmetry is constrained by $|A| < 0.0045$ for $\ell \gtrsim 600$. The conflict between the large and small scale measurements raises a challenge for theoretical paradigm of the early universe. To reconcile this conflict, it was suggested that the amplitude of power asymmetry shall be strongly scale-dependent \citep{Erickcek:2008sm, Erickcek:2008jp, Watanabe:2010fh, Dai:2013kfa, Lyth:2013vha, Liu:2013kea, McDonald:2013aca, Namjoo:2013fka, Liddle:2013czu, Mazumdar:2013yta, Abolhasani:2013vaa, Cai:2013gma, Kohri:2013kqa, Kanno:2013ohv, Firouzjahi:2014mwa, Lyth:2014mga, Namjoo:2014nra, Kobayashi:2015qma, Kothari:2015tqa, Jazayeri:2014nya, Cai:2015xba, Mukherjee:2015wra, Byrnes:2015dub, Byrnes:2016zxb, Byrnes:2016uqw, Yang:2016wlz,Ashoorioon:2016hemispherical, Jazayeri:2017szw}. In general, this power asymmetry could be modelled as a modulation of certain physical parameters that can affect the power spectrum of CMB anisotropies. For instance, in the {\it multi-speed} inflation models \citep{Cai:2009hw, Cai:2008if} the hemispherical power asymmetry can be interpreted as a consequence of the spatially varying modulation of primordial sound-speed parameter during inflation \citep{Cai:2013gma, Cai:2015xba}.

In order to investigate in detail the scale dependence of the hemispherical power asymmetry, a number of studies have been made extensively on the observables of statistical anisotropy for which characteristic scales lie between large and small scales, e.g.\ \cite{Pullen:2010zy, Baghram:2013lxa, Alonso:2014xca, Pitrou:2015iya, Pereira:2015jya, Shiraishi:2015lma, Hassani:2015zat, Zibin:2015ccn, Bengaly:2015xkw, Shiraishi:2016omb, Shiraishi:2016wec,khosravi201821}. \cite{Shiraishi:2016omb} pointed out that the angular power spectrum of the 21-cm fluctuations from the dark ages could open a new window for testing the power asymmetry. The 21-cm line corresponds to the hyperfine transition of atomic hydrogen. Therefore, the density fluctuations of neutral hydrogen can leave imprint on the 21-cm temperature anisotropies \citep{Loeb:2003ya, Bharadwaj:2004nr}. In contrast with the CMB map in which cosmological information is two-dimensional at the last scattering surface, the 21-cm observations can make a three-dimensional mapping which covers most of observable volumes within our horizon\citep{Mao:2008ug,2012JCAP...05..028L}. Consequently, the 21-cm mapping can offer the constraints on various astrophysical and cosmological models \citep{Cooray:2004kt, Pillepich:2006fj, Munoz:2015eqa, Shimabukuro:2015iqa} in the early universe \citep{Lewis:2007kz}. In particular, the tomographic 21-cm mapping from the epoch of reionization (EoR) could place the constraints on cosmological parameters with the unprecedented level of accuracy \citep{Mao:2008ug, Mao:2013yaa}.

In this paper, we employ the power spectrum of 21-cm fluctuations from the EoR to investigate the scale dependence of the amplitude of hemispherical power asymmetry. To describe its scale dependence, we put forward a new parametrization of the asymmetry amplitude as a function of the comoving wavenumber, inspired by a type of the {\it multi-speed} inflation models. Based on this new parameterization, we forecast the constraints from 21-cm cosmology on the asymmetry amplitude at different scales.

Our paper is organized as follows. In Section 2, we introduce a new parametrization form of the amplitude of the scale-dependent power asymmetry, and then model the 21-cm power spectrum during the EoR with the effect arising from this new parametrization. In Section 3, we employ the Fisher matrix formalism to estimate the error of the hemispherical power asymmetry measurements with the 21-cm observations using future interferometer arrays. We make concluding remarks in Section 4.

\section{The 21-cm power spectrum with scale-dependent power asymmetry}

In this section we begin with a brief discussion on the primordial origin of the hemispherical power asymmetry and propose a new parametrization including the scale-dependent effect that is advocated by some inflation models, such as the multi-speed inflation model. We then study the power spectrum of the 21-cm brightness temperature fluctuations during the EoR under this parametrization. 

\subsection{A new parametrization of the scale-dependent power asymmetry}

The hemispherical power asymmetry, as indicated by the WMAP and Planck experiments, can be modelled as the dipolar modulation of density fluctuations with respect to some special direction. The scale-dependent power spectrum of primordial curvature perturbations under this modulation can be written as \citep{Cai:2015xba}
\begin{equation}
 P_{\zeta}(k, {\bf n},z) = P_{\zeta}^{\rm iso}(k,z) \Big[ 1+2A(k) \,{\bf p} \cdot {\bf n} \Big] \,,
\label{eqn:Pzeta}
\end{equation}
where $P_{\zeta}^{\rm iso}(k)$ is the isotropic power spectrum, $A(k)$ denotes the scale-dependent amplitude of dipolar asymmetry, ${\bf p} \cdot {\bf n}$ reflects the dipolar modulation between the line-of-sight (LOS) of the observer (with unit vector ${\bf n}$) and some preferred direction (with unit vector ${\bf p}$).

For a simple version of the multi-speed inflation model, two scalar fields are minimally coupled to Einstein gravity during the inflationary epoch. While the inflaton field is of the K-essence type, the entropy field may be canonical but it can affect the scale dependence of the sound speed parameter for the inflaton \citep{Cai:2013gma}. As a result, a scale-dependent power asymmetry can be induced by a spatial variation of the sound speed parameter at large scales by introducing a long wavelength perturbation from the entropy field \citep{Cai:2015xba} without violating the bound of over large primordial non-Gaussianities \citep{Adhikari:2015yya, Namjoo:2014nra, Namjoo:2013fka}. In particular, for this type of inflation models the amplitude of power asymmetry can be approximated by a power law of $\ln(k)$. For instance, for a type of string theory inspired inflation model \citep{Cai:2015xba}, $A(k) \sim [\ln(k)]^{3/2}$.

In light of the aforementioned relation, we parameterize a {\it generic} scale-dependent form for the amplitude of power asymmetry as follows,
\begin{equation}
\label{eqn:Afk}
 A(k) = A_0 \, \Big[ \frac{ \ln(k/k_{\rm LSS}) }{ \ln(k_{\rm CMB}/k_{\rm LSS}) } \Big]^{n_a}\,,
\end{equation}
for $k<k_{\rm LSS}$. Here we set the pivot scales to be $k_{\rm LSS}=1 ~ {\rm Mpc}^{-1}$ and $k_{\rm CMB}=0.0045 ~ {\rm Mpc}^{-1}$ (corresponding to $\ell$ over $10^5$ and $\ell=64$, respectively). The 21~cm surveys are sensitive to $k\sim 0.1 ~ {\rm Mpc}^{-1}$, which corresponds roughly to $l=600$. Note that, when $k=k_{\rm CMB}$, $A(k)=A_0$, and thus $A_0$ is constrained by the CMB observations such as the Planck data. Moreover, when $k = k_{\rm LSS}$, the power asymmetry automatically vanishes at the LSS scale in our parametrization form. For simplicity, we take a cutoff of the power asymmetry at the length scales smaller than the LSS scale.

\subsection{The 21-cm power spectrum}
\label{sec:21cm}

The 21-cm line is the transition line of atomic hydrogen between two hyperfine states that are split from the ground state due to the spin coupling of the nucleus and electron. The 21-cm brightness temperature from a bulk of hydrogen gas depends on the matter density, the neutral fraction, the velocity gradient and the spin temperature. During the EoR, it is reasonable to assume that X-ray heating and Ly$\alpha$-pumping effects are so efficient that the HI spin temperature is much greater than the CMB temperature. In this limit, the power spectrum of 21-cm brightness temperature fluctuations can be written as \citep{Datta:2011hv, Mao:2013yaa}
\begin{equation}
\label{eqn:21power}
 P_{\Delta T}({\bf k},{\bf n},z) = \widetilde{\delta T}_b^2 \bar{x}_{\rm HI}^2\,\left[ b_{\rho_{\rm HI}}(z) + \mu_{\bf k}^2 \right]^2\, P_{\zeta}(k,{\bf n},z) \,,
\end{equation}
at large scales, with
\begin{equation}
\label{eqn:widetilde_delta_T_b}
 \widetilde{\delta T}_b(z) = (23.88\,{\rm mK}) (\frac{\Omega_b h^2}{0.02}) \sqrt{\frac{0.15}{\Omega_M h^2} \frac{1+z}{10}} ~. 
\end{equation}
Note that $\bar{x}_{\rm HI}$ is the global neutral fraction which implicitly depends on the redshift, and $\mu_{\bf k} \equiv {\bf k} \cdot {\bf n}/ |{\bf k}|$ (cosine of angle between line-of-sight ${\bf n}$ and wave vector ${\bf k}$ of a given Fourier mode). We can relate the neutral hydrogen density fluctuations to the total matter density fluctuations with a simple bias parameter, the {\it neutral density bias}, $b_{\rho_{\rm HI}}(k,z) \equiv \tilde{\delta}_{\rho_{\rm HI}}({\bf k},z)/\tilde{\delta}_{\rho}({\bf k},z)$. At large scales, the bias approaches to a scale-independent value, $b_{\rho_{\rm HI}}(z)$. Here, we assume that the baryonic distribution traces the cold dark matter distribution at large scales, so $\delta_{\rho_{\rm H}} = \delta_{\rho}$. Also, we assume that no additional hemispherical power asymmetry was generated during the evolution of the universe, so the dipolar modulation of the curvature perturbation is transferred to the 21-cm power spectrum with the same form.

The values of global neutral fraction $ \bar{x}_{\rm HI}$ and neutral density bias $b_{\rho_{\rm HI}}(z)$ rely on the modelling of cosmic reionization. We follow the basic methodology in \cite{Mao:2013yaa}. In addition, we have neglected the contribution of primordial non-gaussianities in density fluctuations in our case, and hence $b_{\rho_{\rm HI}}(z)$ is scale-independent. To be specific, we consider two methods as follows.

(i) Excursion set model of reionization (ESMR). 
The ESMR \citep{Furlanetto:2004nh} is an analytical model that can predict the evolution and morphology of reionization. Its basic assumption is that the local ionized fraction within a spherical region with comoving radius $R$ is proportional to the local collapsed fraction of mass in luminous sources if the mass of source is above some threshold $M_{\rm min}$, i.e.\ $x_{\rm HII}(M_{\rm min},R,z) = \zeta_{\rm ESMR}\, f_{\rm coll}(M_{\rm min},R,z)$, where $\zeta_{\rm ESMR}$ is a free parameter characterizing the efficiency of the mass in releasing ionizing photons into the IGM. Here $f_{\rm coll}(M_{\rm min},R,z)$, the local collapsed fraction smoothed on scale $R$, can be calculated using the excursion set theory for halo model.

The morphology of ionized regions can be determined in the ESMR model in terms of the {\it ionized} density bias, $b_{\rho_{\rm HII}}(z) \equiv \tilde{\delta}_{\rho_{\rm HII}}({\bf k},z)/\tilde{\delta}_{\rho}({\bf k},z)$. It is related to the neutral density bias by $b_{\rho_{\rm HI}}=(1-\bar{x}_{\rm HII}\,b_{\rho_{\rm HII}})/\bar{x}_{\rm HI}$. With Gaussian initial condition and spherical collapse model, one finds $b_{\rho_{\rm HII}}=1+ (2/\delta_c)(\partial  \ln \bar{f}_{\rm coll} / \partial  \ln S_{\rm min})$ \citep{DAloisio:2013mgn}, where $\bar{f}_{\rm coll}(M_{\rm min},z)$ is the global collapse fraction, $S_{\rm min}$ is the variance of density fluctuations smoothed on scale $M_{\rm min}$, and $\delta_c$ denotes the critical density in the spherical collapse model.
Since $\bar{x}_{\rm HI}(z)$ and $b_{\rho_{\rm HII}}(z)$ as a function of redshift are set by the simple parameter $\zeta_{\rm ESMR}$ in the ESMR model, the 21-cm power spectrum is determined by three free parameters ($\zeta_{\rm ESMR}$, $A_0$, $n_a$), given a fiducial cosmology and some preferred direction ${\bf p}$.

We note that, under this model, the power spectrum has its own contribution to excursion probability and correspondingly to the ionized fraction. However, since it is already an analytical simplification of the actual reionization process, the contribution from the power asymmetry is trivial and hence has been ignored in the present study.

(ii) Phenomenological (``pheno-'') model: We may treat $\bar{x}_{\rm HI}$ and $b_{\rho_{\rm HII}}$ as free parameters which embrace our ignorance of reionization. These parameters at different redshifts are independent. In this ``pheno-model'', the 21-cm power spectrum is determined by four free parameters ($\bar{x}_{\rm HI}$, $b_{\rho_{\rm HII}}$, $A_0$, $n_a$) at each redshift. We evaluate the fiducial values of the first two parameters using the ESMR model, so that the comparison of the ESMR and pheno-model results is on the same footing.

We should mention that the process of reionization can, in principle, be affected by the power asymmetry in the matter power spectrum. Therefore, there could be a new bias parameter that quantifies the modification of the asymmetric part of the 21-cm brightness temperature power spectrum. However, it is a secondary effect of power asymmetry, so we ignored this bias here for simplicity.

\subsection{The 21-cm observations with radio interferometric arrays}

Radio interferometric arrays can measure the 21-cm brightness temperature from coordinates ${\bf \Theta} \equiv \theta_x \,\hat{e}_x +\theta_y \,\hat{e}_y + \Delta \nu \,{\bf n}$, where $(\theta_x,\theta_y)$ is the angular position on the sky, and $\Delta \nu$ denotes the frequency difference from the redshift $z_*$ at the center of a redshift bin. We adopt the flat sky approximation, which can be satisfied for each patch of sky when the full, large, field of view is divided to many small patches. In each patch, the observed ${\bf \Theta}$-coordinates are related to the physical, 3D Cartesian coordinates, ${\bf r}$ (with origin at the bin center) by $ {\bf\Theta}_\perp={\bf r}_\perp/d_A(z_*)$, and $ \Delta \nu = r_\parallel / y(z_*) $. Here $d_A(z)$ is the comoving angular diameter distance, $y(z) \equiv \lambda_{21}(1+z)^2/H(z)$, where $H(z)$ is the Hubble parameter at $z$, $\lambda_{21}= \lambda(z)/(1+z) \approx 21\,{\rm cm}$.

The interferometric arrays measure the 21-cm power spectrum through the intensities of signals corresponding to the interference patterns of specified baselines, which are different modes in Fourier space. The Fourier dual of the observed ${\bf \Theta}$-coordinates, ${\bf u} \equiv u_x \hat{e}_x + u_y \hat{e}_y  + u_\parallel {\bf n}$ ($u_\parallel$ has the dimension of time), is related to ${\bf k}$ (Fourier dual of ${\bf r}$) by  ${\bf u}_\perp = d_A(z_*)\, {\bf k}_\perp$ and $u_\parallel = y(z_*)\,k_\parallel $. Accordingly, the power spectrum of 21-cm fluctuations in ${\bf u}$-space is given by $ P_{\Delta T}({\bf u},z)= P_{\Delta T}({\bf k},z) /(d_A^2 \,y)$.

\begin{table}
\begin{tabular}{cccc}
Experiment &  $L_{\rm min}$ (m)  & $A_e$($z=6/8/12$)[${\rm m}^2$] & $\Omega$[sr] \\ \hline
SKA        &  51     & 340/560/1170    & $\pi(\lambda/54)^2$\\
Omniscope  &  1      & 1/1/1           & $ 2\pi$\\
\end{tabular}
\caption{Specifications for 21-cm interferometer arrays. We assume the bandwidth $B=6\,{\rm MHz}$ for each redshift bin (\citealt{Mao:2008ug}; also see the SKA technical book).}
\label{tab:spec}
\end{table}

The range of accessible wave vector ${\bf k}$ as well as the noise are determined by array configuration. We consider two upcoming or proposed telescopes herein: the Square Kilometre Array low frequency array (SKA-Low) \footnote{\href{http://www.skatelescope.org/}{\texttt{http://www.skatelescope.org/}}}, and the Omniscope\footnote{\href{http://space.mit.edu/home/tegmark/main_omniscope.html}{\texttt{http://space.mit.edu/home/tegmark/main\_omniscope.html}}} \citep{2009PhRvD..79h3530T,2010PhRvD..82j3501T}. In the Phase 1 of SKA-Low (``SKA1''), the antennae obey the Gaussian distribution. While there are totally $866$ antenna stations in a core region, about $650$ stations are located inside the innermost region of the core (with radius less than $1$ kilometer). The contributions from antennae outside this innermost region are negligible. The design of the Phase 2 of SKA (``SKA2'') has not yet been finalized. We follow the SKA technical book\footnote{\href{https://www.skatelescope.org/key-documents/}{\texttt{https://www.skatelescope.org/key-documents/}}} and assume that the noise of SKA-Low in Phase 2 is four times smaller than that in Phase 1. For Omniscope, we assume that a total of $10^6$ antennae spread out uniformly with occupation fraction close to unity. We list the details of specifications in Table~\ref{tab:spec}.

The noise of 21-cm measurements has two sources: synchrotron radiation emitted by our Galaxy and noise of antenna. The noise in power spectrum in ${\bf u}$-space is given by \citep{McQuinn:2005hk, Mao:2008ug}
\begin{equation}
P_{N}({\bf u}_\perp)=\left( \frac{\lambda T_{\rm sys}}{A_e}\right)^2 \frac{1}{t_0 {\bar n}({\bf L}_{{\bf u}_\perp})}\,,
\end{equation}
where $\lambda$ is the wavelength of the signal, $A_e$ denotes effective collecting area, and $t_0$ is the total observation time.
The system temperature is given by $T_{\rm sys} =280\,{\rm K}\, ((1+z)/7.5)^{2.3}$ \citep{wyithe2007biased}, where $z$ is the redshift. 
For an interferometric array, a baseline ${\bf L}$ corresponds to ${\bf u_\perp} = 2\pi {\bf L}/\lambda$. To reduce the noise, a Fourier mode is measured in multiple times using redundant baselines. The number of redundant baselines, $\bar{n}({\bf L}_{\bf u_\perp})d^2{\bf L}$, for the Fourier mode ${\bf u_\perp}$ corresponding to the baseline ${\bf L}_{\bf u_\perp}$, can be evaluated for a given antenna distribution. The minimum $k_{\perp}$ is determined by the minimum baseline, $k_{\perp,\rm min} = 2\pi L_{\rm min}/(\lambda \,d_A)$, e.g.\ $k_{\perp,\rm min} = 0.0125\, {\rm Mpc}^{-1}$ for the SKA and $0.0002\, {\rm Mpc}^{-1}$ for the Omniscope, at $z=11.24$. The studies of \citep{McQuinn:2005hk, Lidz:2013tra} demonstrated that the foreground contamination can be neglected for $k_\parallel \ge k_{\parallel,{\rm min}}= 2\pi/(yB)$, where $B$ is the bandwidth of the redshift bin, e.g.\ $k_{\parallel,{\rm min}}= 0.055\, {\rm Mpc}^{-1}$ at $z=11.24$. Thus the minimum wavenumber that the 21~cm experiments can probe is $k_{\rm min} = \sqrt{k_{\parallel,{\rm min}}^2 + k_{\perp,\rm min}^2}$, e.g.\ $k_{\rm min} \simeq 0.056\, {\rm Mpc}^{-1}$ at $z=11.24$. Since we make the linear bias assumption which is valid only at large scales, we set the maximum wave number $k_{\rm max} = 0.15\,{\rm Mpc}^{-1}$. 
In principle, 21~cm observations can exploit a much larger number of modes at smaller scales to constrain cosmology, but this requires sophisticated modelling of astrophysical processes involving reionization and gas temperature, because the nonlinear physics kicks in at the small scales. As such, we choose to work in the clean regime for cosmology, $k \le 0.15\,{\rm Mpc}^{-1}$. 
The range of scales that 21-cm observations are sensitive to, $0.056 \lesssim k_{\rm 21cm} \lesssim 0.15 \,{\rm Mpc}^{-1}$, stands between those of the CMB and the LSS, thereby making the 21-cm line an independent probe of the hemispherical power asymmetry.

To estimate the sensitivity of radio interferometric arrays, we employ the Fisher matrix formalism. For a given set of parameters $\{\lambda_a\}$, the Fisher matrix for 21-cm power spectrum can be written as ${\bf F}_{ab}=\sum_{\bf u} \left(\frac{\partial P_{\Delta T}({\bf u})}{\partial \lambda_a}\right) \left(\frac{\partial P_{\Delta T}({\bf u})}{\partial \lambda_b}\right)/[\delta P_{\Delta T}({\bf u})]^2$, where $\delta P_{\Delta T}({\bf u}) =  \left[P_{\Delta T}({\bf u})+P_N(u_\perp)\right]/\sqrt{N_c}$. Here $N_c$ is the number of independent modes in that pixel, $N_c=u_\perp du_\perp du_\parallel \Omega B/(2\pi)^2$, where $\Omega$ is solid angle spanning the field of view. We adopt logarithmic pixelization, $du_\perp/u_\perp = du_\parallel/u_\parallel = 10\%$. The $1\sigma$ forecast error of the parameter $\lambda_a$ is given by $\Delta \lambda_a = \sqrt{({\bf F^{-1}})_{aa}}$. 
Note that due to the power asymmetry, we pixelize the sky in the calculation of Fisher matrix. In a given pixel, the power spectrum is affected by the angle between the LOS of this pixel and the preferred direction, according to Eq.~(\ref{eqn:Pzeta}). We calculate the Fisher matrix from each pixel and add up the contributions of all pixels in the total field of view of a survey. Thus the result of Fisher forecast depends on the direction of the observed sky, unless an observation like the Omniscope covers the whole sky.

Our fiducial cosmology is described as follows: $\Omega_{\Lambda}=0.68$, $\Omega_{\rm M}=0.32$, $\Omega_{\rm b}=0.049$, $H_0 = 100h\,{\rm km}\,{\rm s}^{-1}\,{\rm Mpc}^{-1}$ ($h=0.67$),  $\sigma_8=0.83$, $n_{\rm s}=0.96$, where these parameters are consistent with the Planck 2015 results \citep{Ade:2015xua} and with the matter power spectrum of \cite{Eisenstein:1997jh}.

\section{Results}

The power asymmetry is parametrized with two parameters, $A_0$ (the power asymmetry amplitude) and $n_a$ (spectral tilt characterizing the scale dependence), respectively. In some inflation models, the value of tilt is predicted to some specific value, while the amplitude is free. As such, we first consider the scenario in which the tilt is fixed at $n_a=3/2$, as predicted by the multi-speed inflation model, and forecast the detectability of $A_0$ with 21~cm experiments in this case. This scenario can be treated as the minimal set of power asymmetry model, since non-zero detection of $A_0$ would directly confirm the hemispherical power asymmetry at the scale of relevance to 21~cm experiments. We then investigate the general scenario in which both $A_0$ and $n_a$ are free. The constraint of the tilt would tell the shape, or the scale dependence, of power asymmetry. The fiducial values are taken at $A_0=0.072$,  as indicated by CMB experiments \citep{Ade:2015hxq} and consistent with previous work \citep{Akrami:2014eta, Yang:2016wlz}, and $n_a=3/2$, unless noted to be varied.

For 21~cm observational data, we first consider the 21-cm power spectrum measurement from a single redshift-bin at $z = 11.24$ and at $z = 10.00$, which corresponds to $\bar{x}_{\rm HI} \approx 0.67$ and $0.33$ in our fiducial reionization model, respectively, for illustrative purpose. We then consider the information of 21-cm power spectra from multiple redshift-bin measurements by combining $7$ redshift bins with a total bandwidth of 42~MHz ($z \approx 9.5 - 13.4$).

For the methodology of modelling reionization, we use both the ESMR model and the pheno-model for single redshift constraints. Since in the  pheno-model $\bar{x}_{\rm HI}(z)$ and $b_{\rho_{\rm HII}}(z)$ at different redshift bins are independent parameters, combining 21~cm power spectra from multiple redshift bins cannot improve the sensitivity of power asymmetry parameters with the pheno-model \citep{Mao:2013yaa}. Thus we only employ the ESMR model for multiple redshift-bin measurements. 

We assume the fiducial value of $\zeta_{\rm ESMR}=50.0$, which, in the case of no power asymmetry, corresponds to $\bar{x}_{\rm HI}=0.67$ (0.33) and $b_{\rho_{\rm HII}}=5.89$ (5.12) at $z = 11.24$ (10.00), respectively, and, for multiple redshift-bin measurements, $\bar{x}_{\rm HI} = 0.13-0.91$ for the redshift range $z \approx 9.5 - 13.4$. 

\begin{table}
\centering
\begin{tabular}{cccc|ccc}
 & & \multicolumn{2}{c}{ESMR} & \multicolumn{3}{c}{pheno-model} \\
    \cline{3-4}\cline{5-7} 
    $\bar{x}_{\rm HI}$  & Experiment & $A_0$ & $\zeta_{\rm ESMR}$ & $A_0$ & $ \bar{x}_{\rm HI}$ & $ b_{\rho_{\rm HII}}$ \\
    \hline
    &  $\bigl\lbrack$ F.V. & 0.072 & 50.0 & 0.072 & 0.67 & 5.89 $\bigr\rbrack$ \\
    $0.67$ & SKA2 & 0.14 & 0.38 & 0.35 & 0.056 & 1.23 \\
    & Omniscope  & 0.0053 & 0.0088 & 0.0060 & 0.0010&0.021 \\ 
    \hline
    & $\bigl\lbrack$ F.V. & 0.072 & 50.0 & 0.072 & 0.33 & 5.12 $\bigr\rbrack$ \\ 
    $0.33$ & SKA2 & 0.41 & 2.85 & 0.43 & 0.12&1.09 \\
    & Omniscope  & 0.0054 & 0.0218  &  0.0054 & 0.0036 & 0.031 \\
\end{tabular}
\caption{The $1\sigma$ forecast errors at $\bar{x}_{\rm HI} \approx 0.67$ ($z=11.24$) and $\bar{x}_{\rm HI} \approx 0.33$ ($z=10.00$), respectively. ``F.V.'' means fiducial values. Here we have fixed the tilt of power asymmetry at $n_a=3/2$, and assume the integration time of 1000 hours for the survey.}
\label{tab:a1}
\end{table}

\begin{figure}
\includegraphics[width=0.23\textwidth]{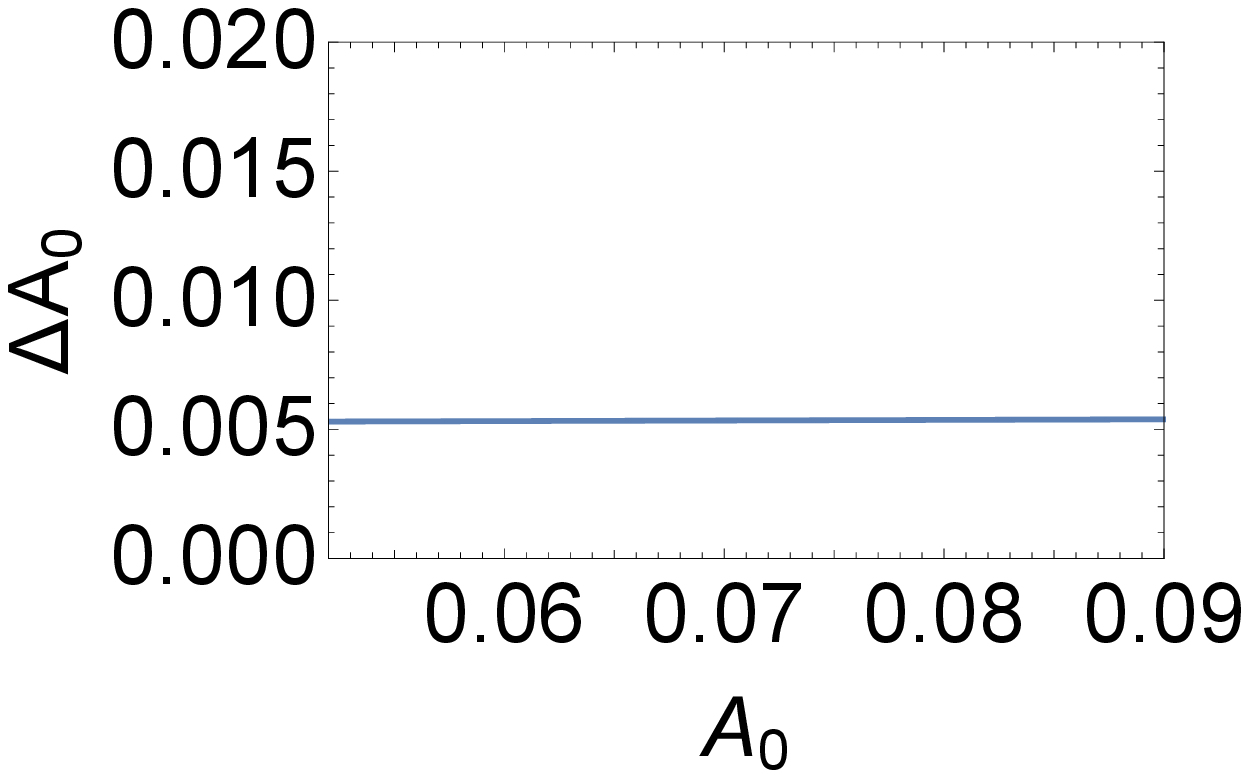}
\includegraphics[width=0.23\textwidth]{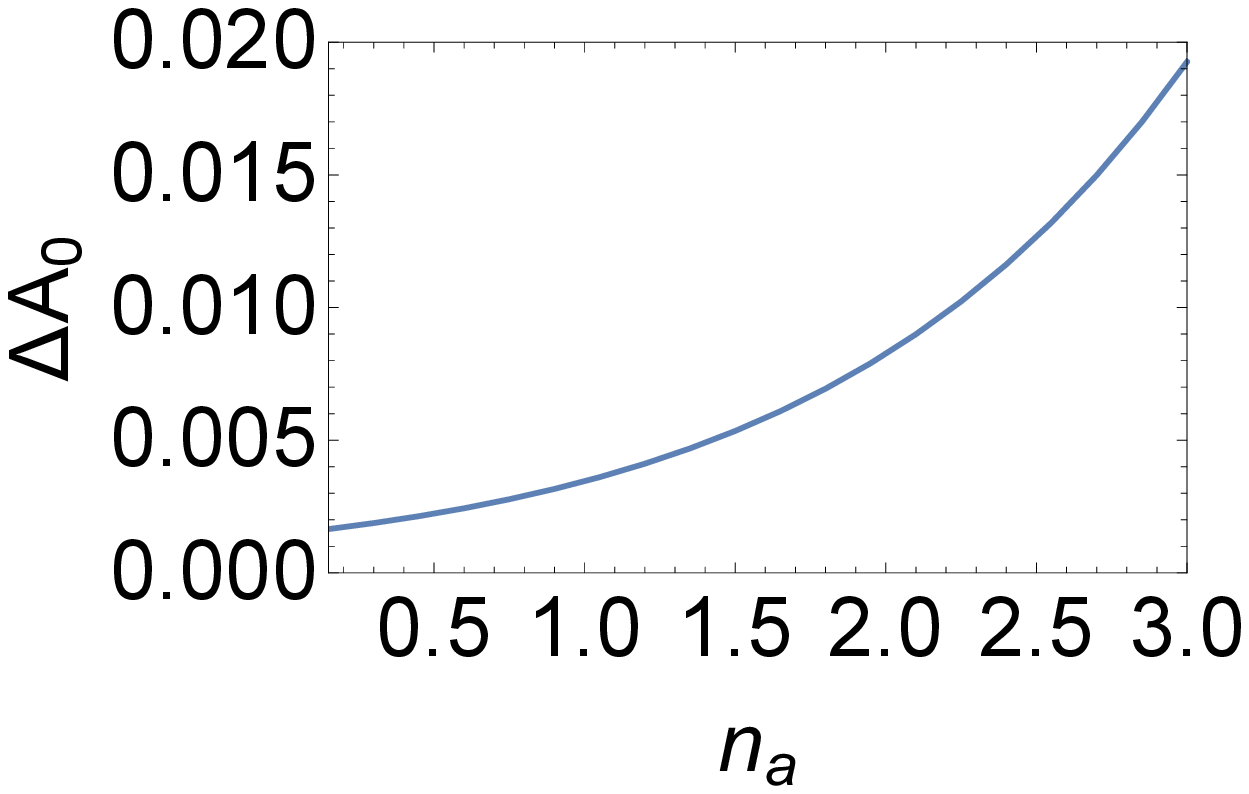}
\caption{Left: the $1\sigma$ forecast error $\Delta A_0$ for the amplitude of power asymmetry $A_0$ as a function of its own fiducial value, when the tilt of power asymmetry is fixed at $n_a=3/2$. 
Right:  $\Delta A_0$ as a function of the fiducial value of $n_a$, when $n_a$ is fixed, and we take the fiducial value $A_0=0.072$.
Here we assume a measurement of a single redshift bin at $z=11.24$ ($\bar{x}_{\rm HI} \approx 0.67$) by the Omniscope with 1000 hours integration time, and employ the ESMR model. }
\label{fig:h0}
\end{figure}

\begin{table*}
\centering
\begin{tabular}{ccccc|ccccc}
	 &   &    \multicolumn{3}{c}{ESMR}         &   \multicolumn{3}{c}{pheno-model}	                \\
	      \cline{3-5}\cline{6-9}
$\bar{x}_{\rm HI}$  & Experiment   & $A_0$ & $n_a$ & $\zeta_{\rm ESMR}$ & $A_0$ & $n_a$ & $ \bar{x}_{\rm HI}$& $ b_{\rho_{\rm HII}}$ \\
	      \hline
     &  $\bigl\lbrack$F.V.    & 0.072	& 1.5   & 50.00   & 0.072       & 1.5   & 0.67     & 5.89 $\bigr\rbrack$  	     \\
$0.67$   & SKA2	     &    0.44& 6.58 & 0.38 &    1.28 & 86.31 & 0.73& 16.11		\\
        & Omniscope  &    0.029 & 0.44 & 0.0088 &   0.030 & 0.46 & 0.0011&0.022	\\ \hline
     & $\bigl\lbrack$F.V.    & 0.072	     & 1.5	 & 50.00     & 0.072   & 1.5   & 0.33 & 5.12 $\bigr\rbrack$ 	    \\ 
$0.33$     & SKA2	     &       0.54 & 14.05 & 5.02 &  1.69 & 111.53& 0.48&8.66  \\
        & Omniscope  &      0.027 & 0.41 & 0.022  &  0.028 & 0.43 & 0.0037&0.032 \\
\end{tabular}
\caption{The same as Table~\ref{tab:a1}, but here both $A_0$ and $n_a$ are free parameters.}
\label{tab:r5}
\end{table*}

\begin{figure*}
\includegraphics[width=0.24\textwidth]{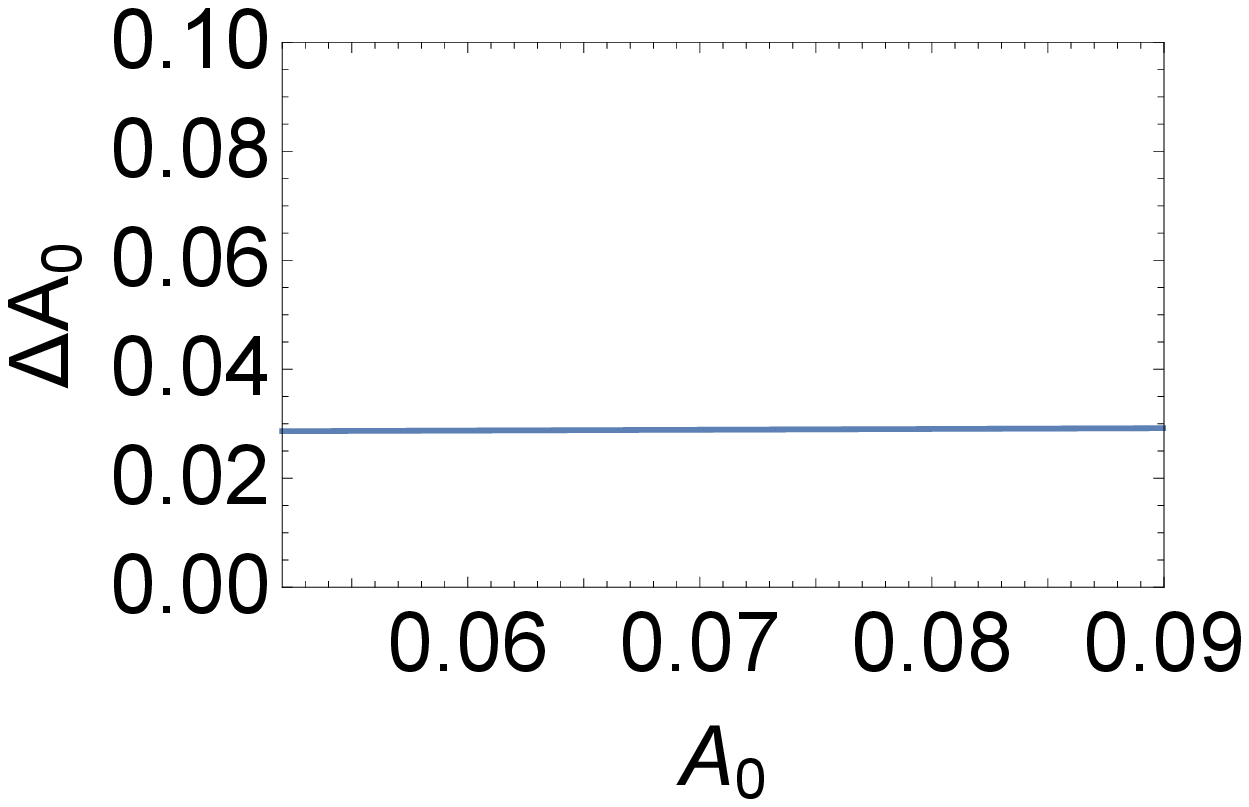}
\includegraphics[width=0.24\textwidth]{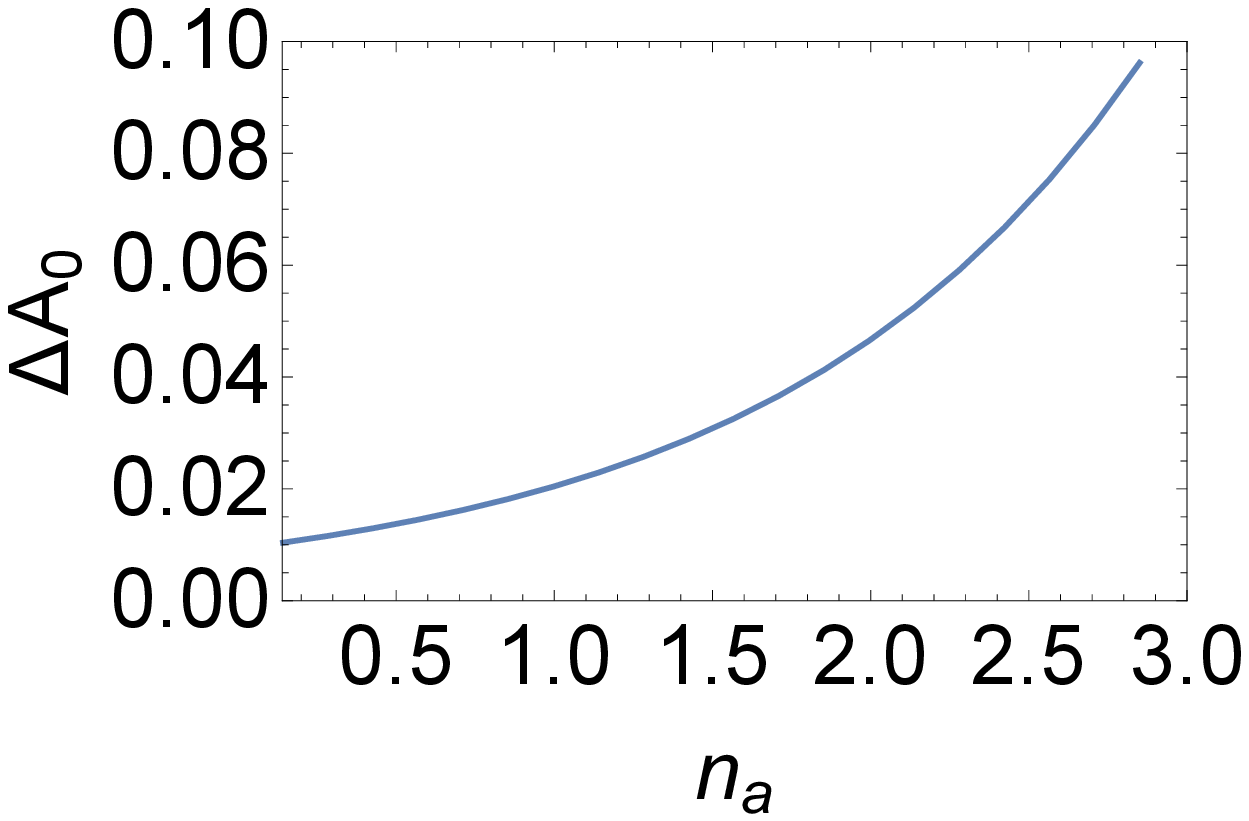}
\includegraphics[width=0.24\textwidth]{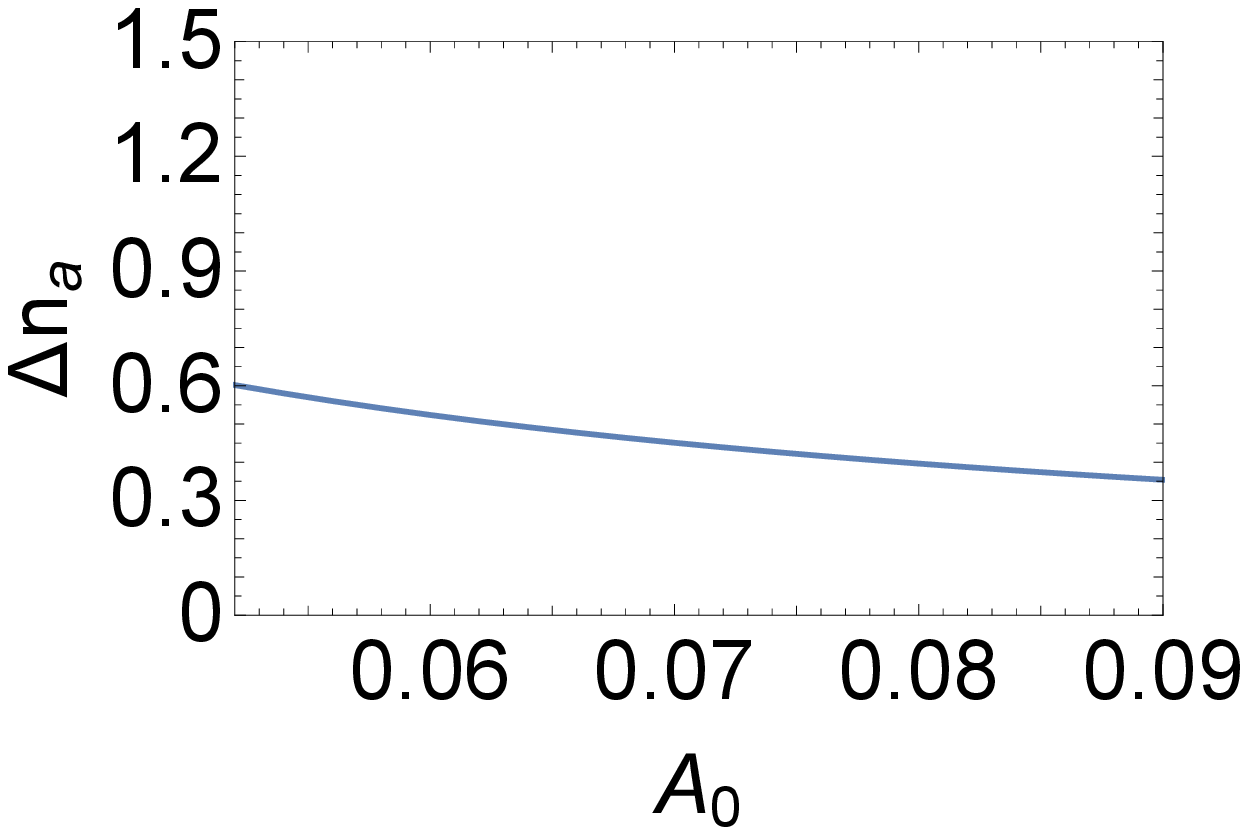}
\includegraphics[width=0.24\textwidth]{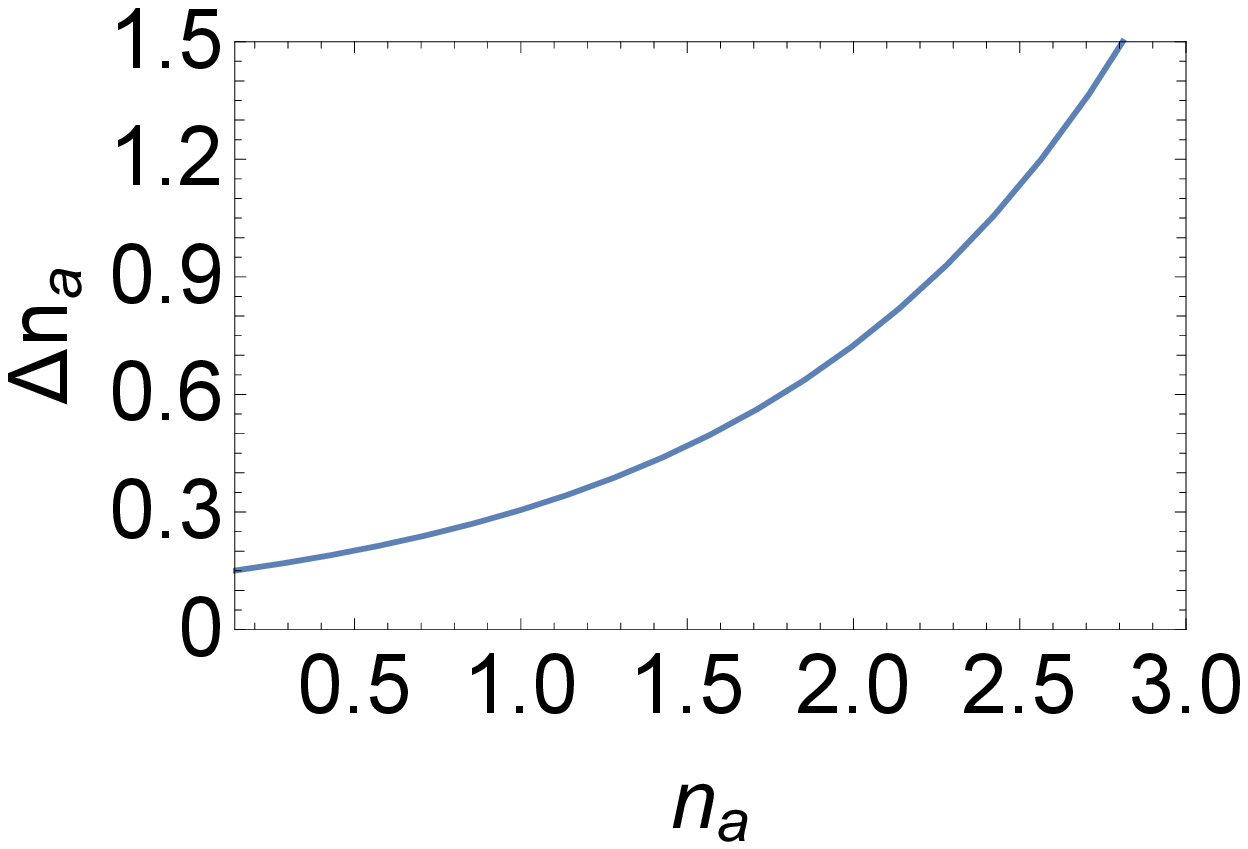}
\caption{From the left to right: the $1\sigma$ forecast error $\Delta A_0$ and $\Delta n_a$ as a function of the fiducial value of $A_0$ and $n_a$, respectively. Here both $A_0$ and $n_a$ are free parameters. Their fiducial values are taken at $A_0=0.072$ and $n_a=3/2$ unless varied. Here we assume a measurement of a single redshift bin at $z=11.24$ ($\bar{x}_{\rm HI} \approx 0.67$) by the Omniscope with 1000 hours integration time, and employ the ESMR model.}
\label{fig:h2}
\end{figure*}

\begin{table*}
\centering
\begin{tabular}{cccc|ccc}
 & & \multicolumn{2}{c}{Fixing $n_a=3/2$} & \multicolumn{3}{c}{Relaxing $n_a$} \\  \cline{3-7}
 Experiment & Integration time & $ A_0$ & $\zeta_{\rm ESMR}$ & $ A_0$ & $n_a$ & $\zeta_{\rm ESMR}$ \\ \hline
$\bigl\lbrack$F.V. &  & 0.072  & 50.00 & 0.072 	& 1.5   & 50.00   $\bigr\rbrack$  \\
 SKA2 & 1000 hours & 0.028 & 0.16 & 0.22 & 3.35 & 0.17 \\
 Omniscope & 1000 hours & 0.0013 & 0.0036 & 0.011 & 0.17 & 0.0037 \\
 Omniscope & 4000 hours & 0.0011 & 0.0025 & 0.0099 & 0.15& 0.0025 \\
\end{tabular}
\caption{The $1\sigma$ forecast errors from the 21~cm power spectra by combining $7$ redshift bins $z \approx 9.5 - 13.4$ ($\bar{x}_{\rm HI} \approx 0.13-0.91$). Here we employ the ESMR model.}
\label{tab:a2}
\end{table*}

\begin{figure}
\centering
\includegraphics[width=.156\textwidth]{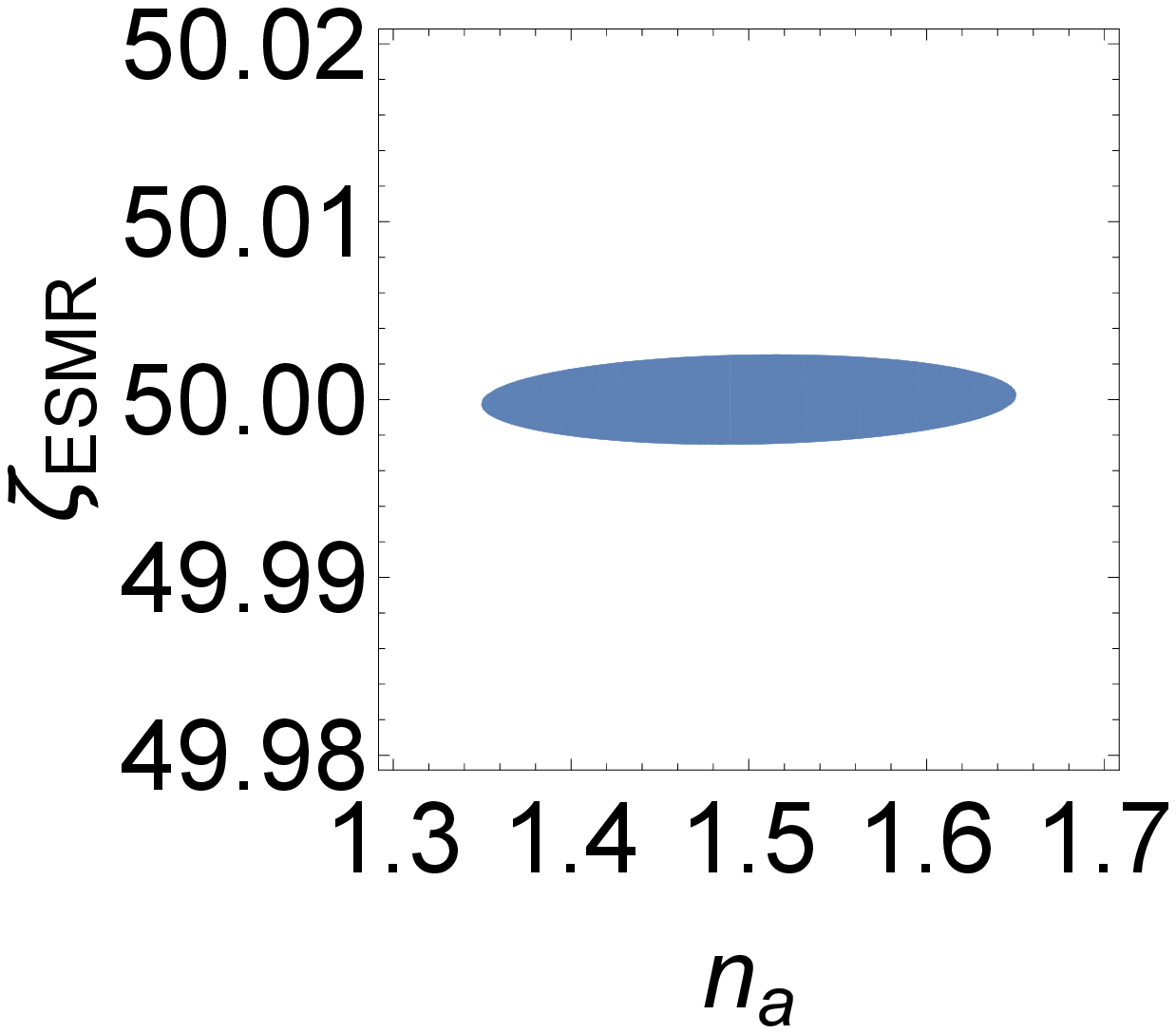}
\includegraphics[width=.15\textwidth]{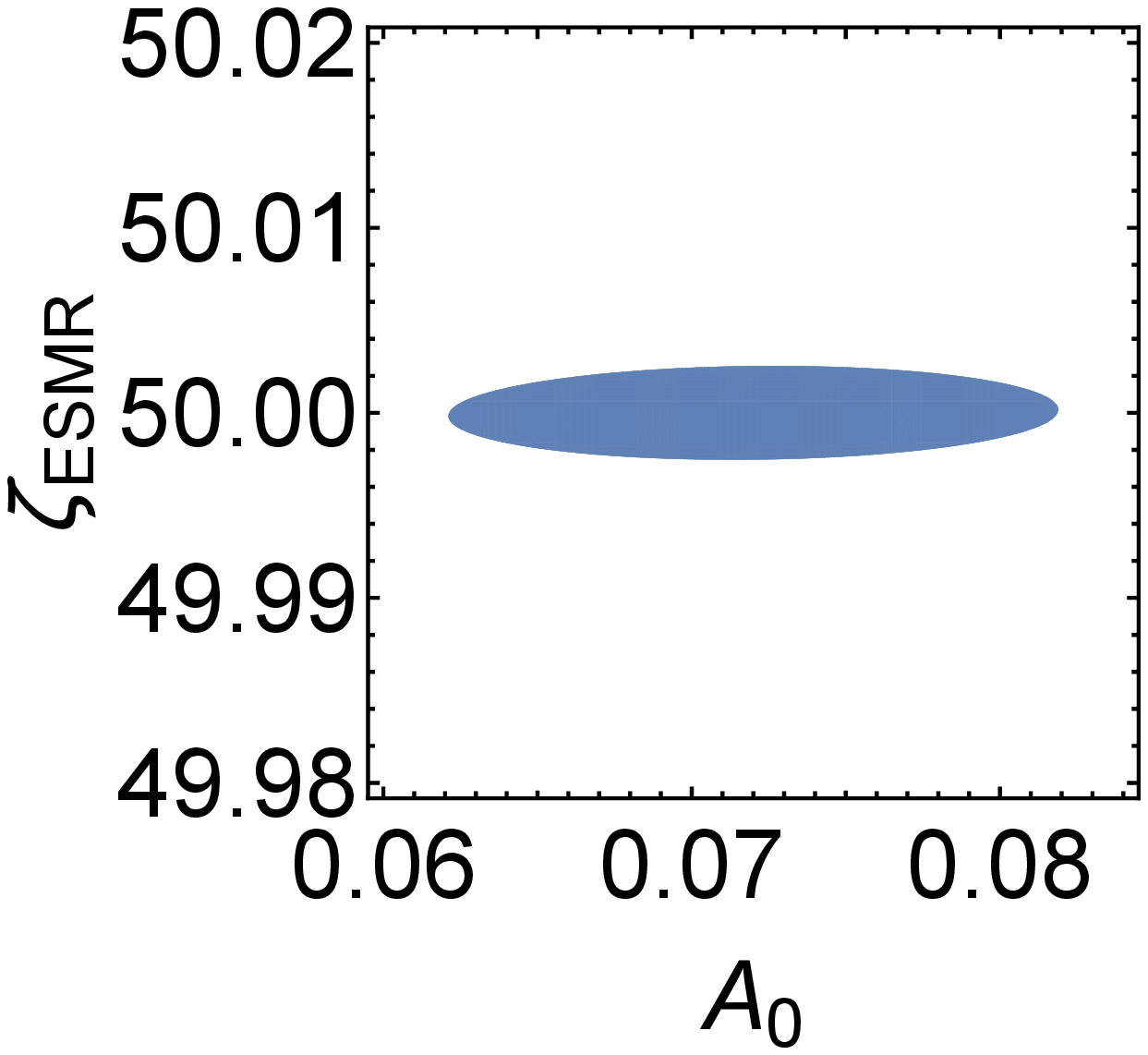}
\includegraphics[width=.135\textwidth]{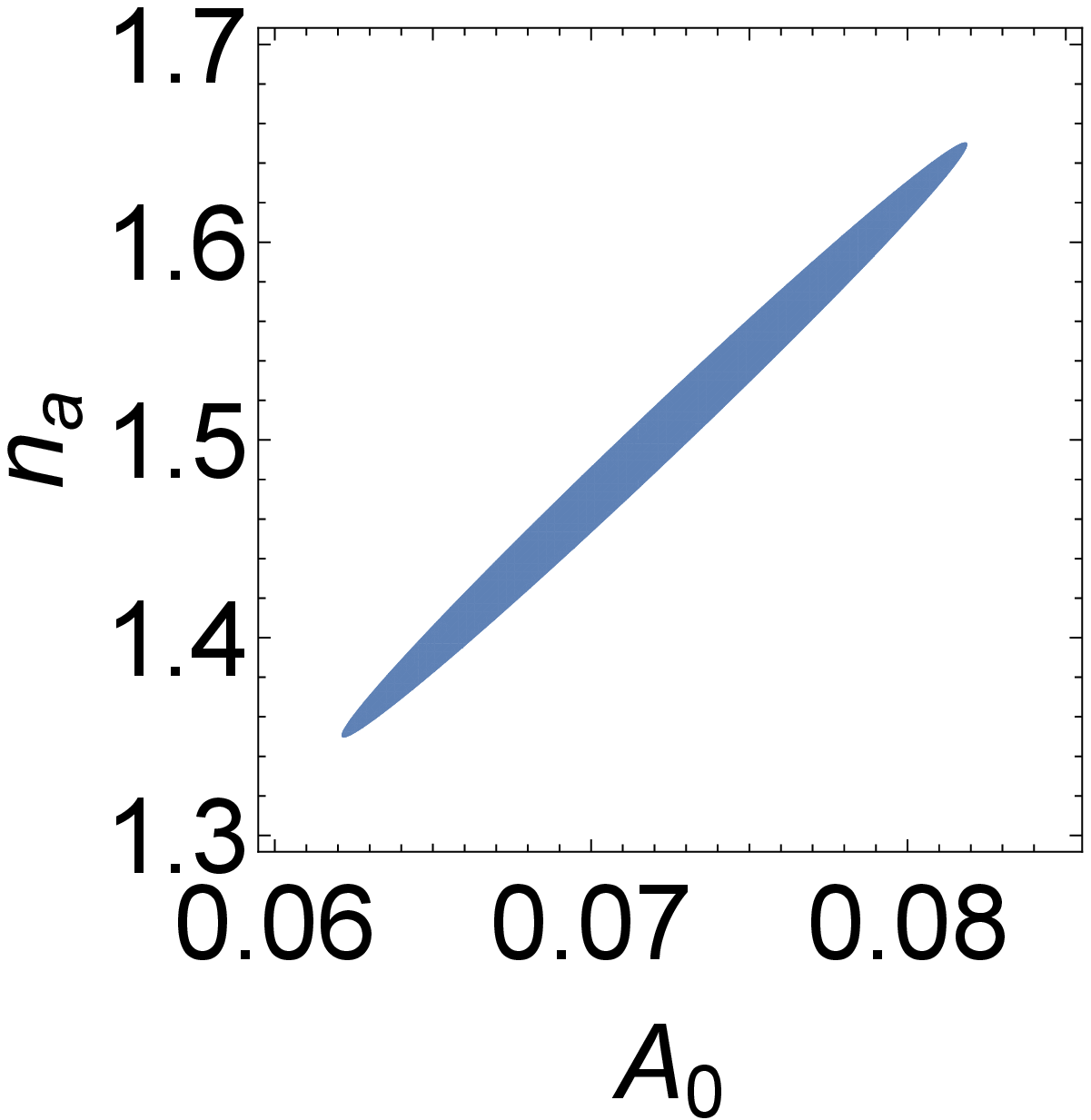}
\caption{Degeneracy of the $1\sigma$ forecast errors between $\zeta_{\rm ESMR}$, $A_0$, and $n_a$, assuming the combination of $7$ redshift-bin measurements ($z \approx 9.5 - 13.4$) from the Omniscope with 4000 hours integration time. The fiducial values are taken at $\zeta_{\rm ESMR}=50.0$, $A_0=0.072$, and $n_a=3/2$.}
\label{fig:contour1}
\end{figure}

\subsection{Constraints from single-redshift measurements}

\subsubsection{Fixing $n_a$}

We first consider the constraints from a single redshift, and fix the tilt of power asymmetry at $n_a=3/2$. Table~\ref{tab:a1} shows the $1\sigma$ forecast error $\Delta A_0$, marginalized over $\zeta_{\rm ESMR}$ in the ESMR model and over $\bar{x}_{\rm HI}$ and $b_{\rho_{\rm HII}}$ in the pheno-model, respectively. We find that $\Delta A_0$ given by these two models are almost identical for the Omniscope, and comparable for SKA2. $\Delta A_0$ by the pheno-model is larger than that by the ESMR model, due to the degeneracy between $A_0$ and $\bar{x}_{\rm HI}$. For this reason, we choose the ESMR as our fiducial reionization model. 

We find that the constraints $\Delta A_0$ from the Omniscope are more than 20 times tighter than given by SKA2. This is due to two reasons. First, the Omniscope has much more redundant baselines for power spectrum measurements, which significantly reduces the thermal noise. Secondly, the Omniscope can observe the whole sky simultaneously, so it exploits the dependence of asymmetric power spectrum on the LOS direction with respect to the preferred direction to break the degeneracy between $A_0$ and $\bar{x}_{\rm HI}$, according to Eqs.~\eqref{eqn:Pzeta} and \eqref{eqn:21power}. For the SKA, nevertheless, we assume that the telescope points to the preferred direction, which we tested gives better constraints than pointing to other directions, and thus such dependence is weakly exploited.

For measurements from different single redshift-bins, i.e.\ centred at $z=11.24$ ($\bar{x}_{\rm HI} = 0.67$) and $z=10.00$ ($\bar{x}_{\rm HI} = 0.33$), respectively, $\Delta A_0$ for the same experiment and same model are comparable. 

In Figure~\ref{fig:h0}, we test the dependence of the forecast error $\Delta A_0$ on the fiducial values we assumed. We find that $\Delta A_0$ is independent of the fiducial value of $A_0$. However, $\Delta A_0$ is sensitive to the fiducial value of $n_a$: when varying $n_a$ from $0$ to $3$, $\Delta A_0$ increases by an order of magnitude. This implies that the detectability of power asymmetry at the intermediate scales by 21~cm power spectrum measurements is independent of its amplitude, but the constraint is sensitive to its scale-dependence $n_a$ and, therefore, the forecast does depend on specific inflation models which give different predictions of $n_a$.

\subsubsection{Relaxing $n_a$}

We now relax $n_a$, so both $A_0$ and $n_a$ are allowed to vary as free parameters. The $1\sigma$ forecast error from single redshift is listed in Table \ref{tab:r5}. We find that for the SKA2, the constraint $\Delta A_0$ when relaxing $n_a$ is comparable to (but larger than) the result when fixing $n_a$. However, for the Omniscope, $\Delta A_0$ is increased 5 times larger by relaxing $n_a$. Since the measurements from the Omniscope is nearly cosmic variance limited, its constraints are more affected by the degeneracy between $\Delta A_0$ and $n_a$. 

For the comparison between different experiments when relaxing $n_a$, we find that the Omniscope can still put constraints on power asymmetry parameters $A_0$ and $n_a$ with accuracies $\Delta A_0$ and $\Delta n_a$ more than 20 times tighter than the SKA2. 

For the comparison between different reionization models, similar to the case of fixing $n_a$, the results using the Omniscope agree very well. For the SKA2, however, $\Delta A_0$ from the pheno-model is 3 times larger than that from the ESMR model, and $\Delta n_a$ from the pheno-model is an order of magnitude larger.

For the comparison between different redshifts, similar to the case of fixing $n_a$, the results for the same experiment and same model are comparable. 

When relaxing $n_a$, in Figure~\ref{fig:h2}, we test the dependence of the forecast errors $\Delta A_0$ and $\Delta n_a$ on their fiducial values. We find that $\Delta A_0$ is independent of the fiducial value of $A_0$. Similarly, $\Delta n_a$ is only weakly dependent of $A_0$. However, both $\Delta A_0$ and $\Delta n_a$ are sensitive to the fiducial value of $n_a$: when varying $n_a$ from $0$ to $3$, both $\Delta A_0$ and $\Delta n_a$ increase by about an order of magnitude. This confirms our previous conclusion that the constraints on power asymmetry are independent (or weakly dependent) of the amplitude, but the errors strongly rely on the scale dependence of power asymmetry and, therefore, on the inflation models.

\subsection{Constraints from multiple-redshift measurements}

We now consider the scenario in which the 21~cm power spectra from 7 redshift bins are combined to constrain the power asymmetry. The redshift range is $z \approx 9.5 - 13.4$, corresponding to the evolution of reionization at $\bar{x}_{\rm HI} \approx 0.13-0.91$. We only employ the ESMR model for the analysis here and show the results in Table~\ref{tab:a2}. When fixing $n_a$, we find that combining 7 redshift bins can improve the constraint on $A_0$ by 5 times tighter than that from a single redshift bin. The constraint on the amplitude given by the SKA2 is $\Delta A_0 = 0.028$, and it is expected that the Omniscope can improve this bound by an order of magnitude. When relaxing $n_a$, the constraints on both $A_0$ and $n_a$ are improved by about 2 times tighter when compared with the single-redshift measurements. This holds for both SKA2 and Omniscope. 

When combining 7 redshift bins, the error of $A_0$ by relaxing $n_a$ is an order of magnitude larger than by fixing $n_a$, for both SKA2 and Omniscope. When comparing the results between different experiments, similar to the constraints from single redshift, the Omniscope can constrain both $A_0$ and $n_a$ about 20 times tighter than the SKA2. While SKA2 has large collection area which results in a high signal-to-noise ratio, the Omniscope measures a baseline several times to reduce its noise as well as its large field of view, enabling it to collect more data.

The Omniscope measurements are nearly cosmic variance limited. To demonstrate this, increasing the integration time 4 times longer (to 4000 hours), which reduces thermal noise, makes almost no difference in the forecast error $\Delta A_0$ and $\Delta n_a$, both for fixing $n_a$ and relaxing $n_a$. 

In Figure~\ref{fig:contour1}, we investigate the degeneracy between $A_0$, $n_a$, and $\zeta_{\rm ESMR}$, when relaxing $n_a$, and show the $1\sigma$ contours. We find that there is no significant degeneracies between $\zeta_{\rm ESMR}$ and $A_0$, and between $\zeta_{\rm ESMR}$ and $n_a$. This is expected, since reionization is not fundamentally affected by the power asymmetry. However, the power asymmetry parameters $A_0$ and $n_a$ are strongly degenerate. This is consistent with the finding that relaxing $n_a$ results in much larger error $\Delta A_0$.

Our upshot is that in the general case, i.e.\ when relaxing $n_a$, a tomographic, multi-frequency, measurements from the Omniscope can put an constraint on $A_0$ with the accuracy $\Delta A \simeq 0.01$. Not only can the multi-frequency measurements by Omniscope realize higher precision in measuring the amplitude of power asymmetry than the current Planck CMB experiment, but also this new constraint will be on different scales of relevance to 21~cm experiment, $0.056 \lesssim k_{\rm 21cm} \lesssim 0.15 \,{\rm Mpc}^{-1}$ in our analysis. An observational constraint on the scale dependence parameter $n_a$ may also be achieved by the Omniscope. We should note, however, that the aforementioned results are strongly dependent on the underlying inflation models that give different predictions of the scale dependence $n_a$ for primordial power asymmetry.

\subsection{Comparison with previous work}

Previously, \cite{Shiraishi:2016omb} suggested that the angular power spectrum of the 21~cm fluctuations from the dark ages could be employed to test the power asymmetry. Similar to our results, they also found that the SKA cannot place better constraints on the amplitude of the power asymmetry than the CMB experiments, but futuristic 21~cm experiments are more promising. In comparison, our work opens a new avenue for testing the power asymmetry with the 21~cm 3D power spectrum from the EOR, in which the 21~cm observations can be technically less difficult than in the dark ages due to less foreground contamination and less thermal noise at lower redshifts. Our work considers not only the case of fixing the tilt $n_a$, but also the general case in which the shape of scale dependence in power asymmetry is allowed to change by varying the tilt, while the shape of scale dependence is fixed in \cite{Shiraishi:2016omb} by their choosing two simpler parameterizations, $A(k)\propto (1-k/k_{\rm LSS})$ or $A(k)\propto (1-k/k_{\rm LSS})^2$.

\section{Conclusions}

We propose a new parametrization of the {\it scale-dependent} hemispherical power asymmetry for the power spectrum of primordial curvature perturbations, inspired by the multiple sound speed inflation model. This parametrization reconciles the conflict between observational constraints from the CMB data on large scales and quasar observations on small scales.

We investigated the 21~cm power spectrum from the EOR with such power asymmetry, based on the excursion set model of reionization as well as the pheno-model. Using the Fisher matrix formalism, we forecast the accuracy with which future 21~cm observations such as the SKA2 and the Omniscope can put constraints on the power asymmetry parameters, $A_0$ (amplitude) and $n_a$ (scale-dependence or the tilt of power asymmetry). 

We find that a multi-frequency measurement with the SKA2 can constrain the amplitude with $\Delta A_0 \simeq 0.2$. This constraint can be improved to $\Delta A_0 \simeq 0.01$ with the Omniscope. Furthermore, if the tilt $n_a$ is fixed, then the constraint on the amplitude can be further improved to $\Delta A_0 \simeq 0.03$ with the SKA2 and $\Delta A_0 \simeq 0.001$ with the Omniscope.

However, even for a multi-frequency measurement with the SKA2, the forecast error on the tilt $\Delta n_a \simeq 3$ is too large to be useful, due to strong degeneracy between $A_0$ and $n_a$. In the long run, the Omniscope can offer tighter constraints on the tilt $\Delta n_a \simeq 0.2$, which will provide supports for the scale dependence of power asymmetry, with a caveat that the error may depend on the fiducial value of $n_a$. 

The intermediate scales at which 21~cm power spectrum measurements can serve as a new window to test the power asymmetry of the Universe is at $0.056 \lesssim k_{\rm 21cm} \lesssim 0.15 \,{\rm Mpc}^{-1}$, complementary to the larger scale probed by the CMB and the smaller scale probed by the LSS. Such measurements will help further constrain inflation models through their predictions of power asymmetry.

\section*{ACKNOWLEDGMENTS}

We thank Xuelei Chen, Dong-Gang Wang and Bin Yue for stimulating discussions. 
BL and YFC wish to thank the Tsinghua Center for Astrophysics for kind hospitality.
BL and ZC are supported in part by the NSFC Fund for Fostering Talents in Basic Science (No. J1310021).
YFC is supported in part by the National Thousand Youth Talents Program of China, by the NSFC (No. 11722327, 11653002, 11421303), by CAST Young Elite Scientists Sponsorship Program (2016QNRC001), and by the Fundamental Research Funds for the Central Universities.
YM is supported in part by the National Key Basic Research and Development Program of China (Grant No.2018YFA0404502, No.2017YFB0203302), 
by the NSFC (Grant No.11673014, 11761141012, 11821303, 11543006), and by the Chinese National Thousand Youth Talents Program. 
Part of numerical simulations are operated on the computer clusters Linda \& Judy in the particle cosmology group at USTC.

\bibliographystyle{mnras}
\bibliography{references}

\bsp   
\label{lastpage}
\end{document}